# Preferences for efficiency, rather than preferences for morality, drive cooperation in the one-shot Stag-Hunt Game

Valerio Capraro[1]  •  Ismael Rodriguez-Lara[2]  •  Maria J. Ruiz-Martos[2]

[1] Department of Economics, Middlesex University London, [2] Department of Economics, Universidad de Granada

August 7, 2019

**Abstract**

Recent work highlights that cooperation in the one-shot Prisoner's dilemma (PD) is primarily driven by moral preferences for doing the right thing, rather than social preferences for equity or efficiency. By contrast, little is known on what motivates cooperation in the Stag-Hunt Game (SHG). Cooperation in the SHG fundamentally differs from cooperation in the PD in that it is not *costly*, but *risky*: players have no temptation to deviate from the cooperative outcome, but cooperation only pays off if the other player cooperates. Here, we provide data from a large (N=436), pre-registered, experiment. Contrary to what has been observed for the PD, we find that SHG cooperation is primarily driven by preferences for efficiency, rather than preferences for doing the right thing.

*Keywords:* morality, cooperation, efficiency, risky choices, stag-hunt game.

# 1. Introduction

There is wide consensus among scholars that people's capacity to cooperate is what has made human societies extremely successful, compared to other animal societies (Trivers, 1971; Ostrom, 2000; Skyrms, 2004; Rand & Nowak, 2013). Psychologists even argue that the psychological basis of cooperation, *shared intentionality*, is what makes humans uniquely humans, as it is possessed by children, but not by great apes (Tomasello et al., 2005). Not surprisingly, a great deal of research has sought to understand what makes people cooperate.

Behavioral economists usually address this question by gathering experimental data on cooperative behavior. Cooperation between two players is typically modeled using either the Prisoner's dilemma (PD) or the Stag-Hunt game (SHG). In the PD, players can either defect (A) or cooperate (B). The payoff consequences of each action are shown in Table 1:

**Table 1. Payoff table in the PD**

|   | A | B |
|---|---|---|
| A | P, P | T, S |
| B | S, T | R, R |

where $T > R > P > S$. Thus, mutual defection (A, A) is the only equilibrium of the one-shot, anonymous, PD. Yet, this gives a smaller payoff than mutual cooperation (B, B), which requires that players override the *temptation* to free-ride by giving up their maximal individual payoff ($T > R$). Hence, PD cooperation is *efficient* but *costly*.

The SHG fundamentally differs from the PD in that people have no temptation to defect, when the other cooperates. Yet, cooperation is *risky*, in that it is beneficial only if both players cooperate. Thus, players in the SHG receive a small but certain payoff if they choose the "safe" option A ($T = P$). Cooperation (i.e., playing B) is "risky" because it offers a greater payoff, but only if the other cooperates. The consequence of this payoff structure ($R > T = P > S$) is that the SHG has two pure Nash equilibria: a risk-dominant equilibrium (A, A) where agents play safe, and a cooperative (payoff-dominant) equilibrium (B, B), which is efficient (Harsanyi & Selten, 1988; Schmidt et al. 2003).

Previous work on what motivates people to cooperate has mainly focused on the PD. There exists overwhelming evidence that people cooperate in this game (Rapoport et al. 1965; Mengel, 2017). This has yielded researchers to consider that people might exhibit distributional, social preferences for equity and/or efficiency (Bolton & Ockenfels, 2000; Engelmann & Strobel, 2004).[1] However, this conclusion has been recently challenged by Capraro & Rand (2018), who found that cooperation in the one-shot, anonymous PD is not primarily driven by distributional motives, but rather by moral preferences for doing the right thing.

By contrast, little is known on what motivates SHG cooperation. This is an important gap as the SHG has served to model cooperative behavior in several settings (Hardin, 1968; Kiss et al. 2017), to the point that it is considered the prototype of the social contract (Rousseau, 1754/1999; Skyrms 2004). Is SHG cooperation, like PD cooperation, driven by non-distributional, moral preferences for doing the right thing? Or, alternatively, is it primarily driven by distributional, social preferences for equity or efficiency? This study follows Capraro & Rand's (2018) methodology to answer these questions.

## 2. Methods

### 2.1. Experimental design and procedures

We conduct our experiment on Amazon's Mechanical Turk (Arechar et al. 2018). The hypotheses, design, sampling and analysis plan were preregistered at http://aspredicted.org/blind.php?x=cv5ja2.[2]

First, participants were randomly matched in pairs to play the SHG in Table 2.[3]

**Table 2. Payoff table in the SHG**

|   | A | B |
|---|---|---|
| A | 25, 25 | 25, 5 |
| B | 5, 25 | 40, 40 |

After making their choices, participants were matched in trios to play the Trade-Off Game (TOG) (Capraro & Rand, 2018; Tappin & Capraro, 2018). In this game, participants had to choose (as dictators) between an "equitable" allocation (13,13,13) that paid the same amount to all members of the trio, and an "efficient" allocation (13, 23,13) that paid more to one of the members, different from the dictator. The TOG choices were framed differently depending on the treatment condition (between-subjects, random assignment):

---

[1] Note that mutual cooperation is more equitable and efficient (under the additional assumption that 2R>T+S) than unilateral defection, thus people might prefer cooperation over unilateral defection if they have social preferences.
[2] See the Appendix for further details, including verbatim instructions.
[3] We adapt Game 2 in Rydval & Ortmann (2005) after dividing the payoffs by 2 to produce roughly 50% of efficient choices in the SHG. A payoff of 25 in our game corresponds to $0.25.

- **TOG efficient frame.** The efficient allocation was labeled as "be generous"; the equitable allocation, as "be ungenerous".
- **TOG equitable frame.** The equitable allocation was labeled as "be fair"; the efficient allocation, as "be unfair".

By this design choice, the efficient (equitable) allocation was framed as morally appropriate in the TOG efficient (equitable) frame, respectively. [4]

Our experiment concludes with standard demographic questions. Participants received a bonus for their choices in the SHG and the TOG in addition to their participation fee ($0.50).

## 2.2. Hypotheses

Capraro & Rand (2018) find that PD cooperation is correlated with the TOG positively framed option (whichever that is), thus they conclude that moral preferences drive PD cooperation. We use the same methodology to contrast the following pre-registered hypotheses:

*Efficiency preferences hypothesis*. SHG cooperation is primarily driven by social preferences for efficiency: there is a positive correlation between the TOG efficient choice and SHG cooperation.

*Moral preferences hypothesis*. SHG cooperation is primarily driven by moral preferences for doing the right thing: there is a positive correlation between the TOG positively framed choice and SHG cooperation.

## 3. Results

We have N=436 participants.[5] Their behavior is summarized in Table 3.[6]

---

[4] Capraro & Rand (2018) show that participants in the TOG efficient (equitable) frame consider to "be generous" ("be fair") as the morally right thing to do.
[5] Following our preregistered protocol, we exclude from the analysis participants who did not answer all the comprehension questions correctly and duplicate responses.
[6] In line with Capraro & Rand (2018) and Tappin & Capraro (2018), we find a strong framing effect: participants are more likely to choose the TOG efficient allocation in the efficient frame (82.73% vs 37.50%). See Appendix C for details on the statistical analysis.

**Table 3. Likelihood of choosing the efficient choice in the SHG and the TOG**

|  | N | SHG cooperation | TOG efficient | SHG cooperation | TOG efficient allocation | SHG cooperation | TOG equitable allocation |
|---|---|---|---|---|---|
| TOG efficient frame | 220 | 61.36 % | 82.73% | 60.99 % | 63.16 % |
| TOG equitable frame | 216 | 64.35 % | 37.50% | 72.84 % | 59.26 % |

As pre-registered, we contrast our hypotheses using logistic regressions to predict SHG behavior (1= cooperation, 0= safe choice). The independent variables include dummy variables for the TOG efficient choice, the TOG efficient frame, and their interaction. Table 4 presents the results (see Appendix C for the marginal effects).

**Table 4. Logit regressions predicting SHG cooperation.**

|  | (1) | (2) |
|---|---|---|
| Constant ($b_0$) | 0.375** | 0.154 |
|  | (0.175) | (0.567) |
| TOG Efficient Frame ($b_1$) | 0.164 | 0.171 |
|  | (0.380) | (0.394) |
| TOG Efficient Choice ($b_2$) | 0.612** | 0.642** |
|  | (0.305) | (0.309) |
| TOG Efficient Frame x TOG Efficient Choice ($b_3$) | -0.704 | -0.801 |
|  | (0.479) | (0.492) |
| Controls (Gender, Age, Education) | No | Yes |
| $\chi_1^2$-test (H$_0$: $b_1 + b_3 = 0$) | 3.40* | 4.54** |
| $\chi_1^2$-test (H$_0$: $b_2 + b_3 = 0$) | 0.06 | 0.17 |
| LR-chi2 | 4.63 | 10.45 |
| Observations | 436 | 426 |

*Note:* Significant at *** p<0.01, ** p<0.05, * p<0.1 Robust standard errors are reported in parenthesis.

In line with the "*efficiency preferences hypothesis*", column (1) reports a positive correlation between the TOG efficient allocation and SHG cooperation. This finding is robust when controlling for gender, age and the level of education, as pre-registered (see column (2)). Interestingly, the results seem to be driven by participants in the TOG equitable frame ($p < 0.046$); by contrast, those choosing the efficient allocation in the TOG efficient frame do not cooperate more than those choosing the equitable allocation ($p > 0.80$). Along these lines, SHG cooperation appears to be (weakly) more frequent among those choosing the TOG efficient allocation in the TOG equitable frame, compared to those choosing the same allocation in the efficient frame ($p < 0.068$).

# 4. Conclusion

Recent work highlights that (one-shot, anonymous) PD cooperation is primarily driven by moral preferences for doing the right thing, rather than by distributional preferences for equity or efficiency (Capraro & Rand, 2018). The current paper adapts the research methodology in Capraro & Rand (2018) to investigate behavior in the SHG, where cooperation is not *costly* (there is no temptation to deviate from the cooperative outcome if the other player cooperates) but it is *risky* (cooperation pays off *iff* the other player cooperates).

Our findings demonstrate that SHG cooperation is primarily driven by efficiency motives, rather than by moral preferences. Interestingly, this effect seems to be driven by the behavior in the TOG equitable frame, where the efficient allocation is negatively framed. We argue that participants choosing the efficient allocation in this frame are likely to have strong preferences for efficiency, strong enough to overcome the moral framing. In fact, people choosing the efficient allocation in the TOG equitable frame might have a stronger preference for efficiency than those in the TOG efficient frame, where the efficient allocation is "nudged".

# References

Arechar, A. A., Gächter, S., & Molleman, L. (2018). Conducting interactive experiments online. *Experimental Economics*, *21*, 99-131.

Bolton, G. E., & Ockenfels, A. (2000). ERC: A theory of equity, reciprocity, and competition. *The American Economic Review*, 90, 166-193.

Capraro, V., & Rand, D. G. (2018). Do the Right Thing: Experimental evidence that preferences for moral behavior, rather than equity or efficiency per se, drive human prosociality. *Judgment and Decision Making*, 13, 99-111.

Engelmann, D., & Strobel, M. (2004). Inequality aversion, efficiency, and maximin preferences in simple distribution experiments. *The American Economic Review*, 94, 857-869.

Hardin, G. (1968). The tragedy of the commons. *Science*, 162, 1243-1248.

Harsanyi, J.C., & Selten, R. (1988). *A General Theory of Equilibrium Selection in Games*. MIT Press, Cambridge

Kiss, H. J., Rodríguez-Lara, I., & Rosa-García, A. (2017). Overthrowing the dictator: A game-theoretic approach to revolutions and media. *Social Choice and Welfare*, 49(2), 329-355.

Mengel, F. (2017). Risk and temptation: A meta-study on prisoner's dilemma games. *The Economic Journal*, 128(616), 3182-3209.

Ostrom, E. (2000). Collective action and the evolution of social norms. *Journal of Economic Perspectives*, 14, 137-158.

Rand, D. G., & Nowak, M. A. (2013). Human cooperation. *Trends in Cognitive Science*, 17, 413-425.

Rapoport, A., Chammah, A. M., & Orwant, C. J. (1965). *Prisoner's dilemma: A study in conflict and cooperation*. Vol. 165. University of Michigan Press.

Rousseau, J. J. (1754). *Discourse on the origin on inequality*. Oxford University Press (1999).

Rydval, O., & Ortmann, A. (2005). Loss avoidance as selection principle: evidence from simple stag-hunt games. *Economics Letters*, 88, 101-107.

Schmidt, D., Shupp, R., Walker, J. M., & Ostrom, E. (2003). Playing safe in coordination games: the roles of risk dominance, payoff dominance, and history of play. *Games and Economic Behavior*, 42, 281-299.

Skyrms, B. (2004). *The Stag Hunt and the Evolution of Social Structure.* Cambridge University Press, Cambridge.

Tappin, B. M., & Capraro, V. (2018). Doing good vs. avoiding bad in prosocial choice: A refined test and extension of the morality preference hypothesis. *Journal of Experimental Social Psychology*, 79, 64-70.

Tomasello, M., Carpenter, M., Call, J., Behne, T., & Moll, H. (2005). Understanding and sharing intentions: The origins of cultural cognition. *Behavioral and Brain Science*, 28, 721-727.

Trivers, R. L. (1971). The evolution of reciprocal altruism. *The Quarterly Review of Biology*, 46, 35-57.